\documentclass[reprint, amsmath, amssymb, groupaddress, aps, prb]{revtex4-1}
\usepackage{graphicx}
\usepackage{amsmath}
\begin{document}
\title{PtP$_2$: An Example of Exploring the Hidden Cairo Tessellation in the Pyrite Structure for Discovering Novel Two-Dimensional Materials}
\author{Lei Liu}
\author{Houlong L. Zhuang}
\email{zhuanghl@asu.edu}
\affiliation{School for Engineering of Matter Transport and Energy, Arizona State University, Tempe, AZ 85287, USA}
\date{\today}
\begin{abstract}
The Cairo tessellation refers to a pattern of type 2 pentagons that can pave an infinite plane without creating a gap or overlap. We reveal the hidden, layered Cairo tessellations in the pyrite structure with a general chemical formula of $AB_2$ and space group $pa\bar3$. We use this hidden tessellation along with density functional theory calculations to examine the possibility of obtaining a two-dimensional (2D) material with the Cairo tessellation from the bulk, using PtP$_2$ as an example. Unlike previously reported single-layer materials such as PdSe$_2$ with a buckled, pentagonal structure\textemdash strictly speaking, not belonging to the Cairo tessellation, we find that single-layer PtP$_2$ is completely planar exhibiting dynamically stable phonon modes. We also observe a reduction in the bandgaps PtP$_2$ from bulk to single layer using the Heyd-Scuseria-Ernzerhof (HSE) hybrid density functional, and the bandgap type switches from indirect to direct. By contrast, using the standard Perdew-Burke-Ernzerhof functional leads to the conclusion of single-layer PtP$_2$ being metallic. We further study the bonding characteristics of this novel single-layer material by computing the Bader charge transfer, the electron localization function, and the crystal orbit overlap population, which show mixed P-P covalent bonding and Pt-P ionic bonding, with the former being stronger. Finally, we study the surface states of single-layer PtP$_2$ and consider the spin-orbit coupling.  We observe no spin-helical Dirac cone states, therefore ruling out single-layer PtP$_2$ as a topological insulator. We expect the example demonstrated in this work will stimulate interest in computationally identifying novel 2D materials from a variety of bulk materials with the pyrite structure.     
\end{abstract}
\maketitle
\section{Introduction}
Hexagon is arguably the most favorable geometry adopted by a number of existing two-dimensional (2D) materials such as single-layer graphene,\cite{novoselov2004electric} boron nitride,\cite{song2010large} molybdenum sulphide,\cite{mak2010atomically} and chromium triiodide\cite{huang2017layer} that exhibit exotic electrical and magnetic properties. As a result of this popularity, a number of 2D materials predicted by density functional theory (DFT) calculations are assumed to adopt hexagonal structures.\cite{zhuang2014computational2, wang2018layer} 

Pentagon, despite its equal simplicity and beauty, had been a headache for a crystallographer who prefers structures with translational periodicity. However, Shechtman {\it et al.} accidentally came upon an Al-Mn alloy with a pentagonal structure.\cite{PhysRevLett.53.1951} This alloy is still called a crystal but with a prefix ``quasi" because of its ``long-range orientational order and no translational symmetry."

To remove this unpleasant prefix, we recently spent efforts in coupling pentagonal geometries with density functional theory (DFT) calculations to predict new 2D crystalline materials. We started with placing atoms of an element at the vertices of the newly discovered type 15 pentagons. We tested eight elements, but we found that no element can form a nanosheet of type 15 pentagons after DFT geometry optimizations.\cite{liu2018can} We then focused on using only one element, carbon, and locate its atoms at the vertices of the other 14 types of pentagons.\cite{liu22018} We found that the carbon nanosheet made of types 2 pentagon remained the same as the initial input type of pentagonal structure. Interestingly, the carbon nanosheet based on type 4 pentagon was optimized into the same structure as resulted from type 2 pentagon. By contrast, the carbon nanosheets built upon the other 12 types of pentagons cannot retain their pentagonal structures. These previous calculations indicate that type 2 pentagon is the most promising pentagon that can be used to discover new 2D materials. 

\begin{figure}
 \includegraphics[width=8cm]{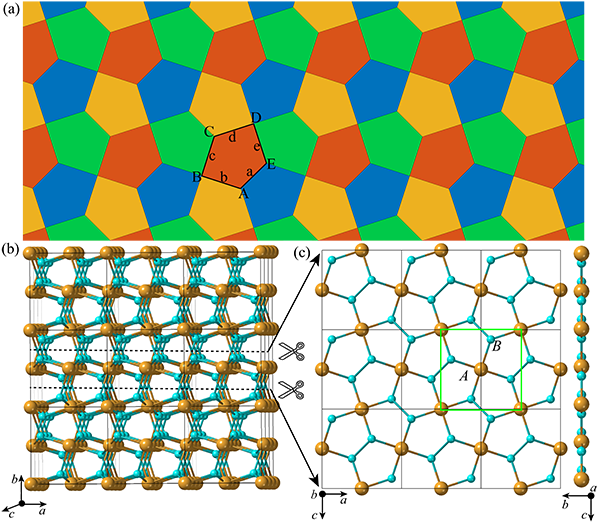}
  \caption{(a) Illustration of the Cairo tessellation formed from type 2 pentagons tiling the plane. The angles and side lengths of a type 2 pentagon are shown. (b) Side view of a 3 $\times$ 3 $\times$ 3 supercell of bulk $AB_2$ with the pyrite structure. (c) Left: Top view of single-layer $AB_2$ extracted from the bulk. Right: Side view of single-layer Pt$P_2$  after DFT geometry optimizations, adopting the Cairo Tessellation. The solid green lines enclose a unit cell of single-layer PtP$_2$ for the calculations.}
  \label{fig:structure}
\end{figure}

Type 2 pentagon is one of the existing 15 types of irregular, convex pentagons discovered so far that can tile a plane without rendering any overlap or gap.\cite{rao2017exhaustive} The topology of this type of pentagon is not unique, because there are only two constraints on the side lengths c and e (c = e) and on the angles B and D (B + D = 180$^\circ$), leaving three degrees of freedom. The pentagon illustrated in Fig.\ref{fig:structure}(a), is a special topology of type 2 pentagon with one two additional geometry constraints: b = c = d = e and B = D = 90$^\circ$. The tiling of this special topology leads to the so-called Cairo tessellation, a pattern that gains its name, as it is ubiquitous on the streets of Cairo in Egypt.\cite{wells1991penguin}

Pentagonal structures with the Cairo tessellation can be straightforwardly identified in van der Waals layered materials such as PdSe$_2$. Each layer of PdSe$_2$ consists of Pd and Se atoms located at the vertices of type 2 pentagons. Note that these atoms are not settled in the same plane. As a result of the weak van der Waals forces, single-layer PdSe$_2$ has been successfully exfoliated from its bulk counterpart with the mechanical exfoliation method, exhibiting a bandgap of 1.3 eV.\cite{oyedele2017pdse2} There are a growing list of single-layer pentagonal materials such as AgN$_3$,\cite{schmidt2007crystal, yagmurcukardes2015pentagonal}SiC$_2$,\cite{lopez2015sigma} CN$_2$,\cite{zhang2016beyond}, B$_2$C\cite{li2015flexible} recently predicted with the buckled Cairo tessellation. Notably, Yang {\it et al.} first reported that single-layer PtN$_2$ exhibits the ideal Cairo tessellation with completely coplanar Pt and N atoms.\cite{liu2018two} 

We believe the above list of single-layer materials with the Cairo tessellation is incomplete. One question naturally arises: How can we search for single-layer materials with the Cairo tessellation? To find the answer to this question, we noticed that the nature has already had abundant compounds\textemdash specially the ones with the pyrite structure \textemdash where the Cairo tessellation is lurking. The pyrite structure possessed by FeS$_2$ is cubic with the general chemical formula of $AB_2$ and space group $pa \bar 3$ (No. 205). A side view of the pyrite structure is displayed in Fig.\ref{fig:structure}(b), showing that atomic layers containing $A$ and $B$ atoms stack along the $b$ direction. The top view of each $AB_2$ layer is shown in Fig.\ref{fig:structure}(c), revealing the hidden Cairo tessellation. We assume that obtaining single-layer $AB_2$ with the Cairo tessellation is analogous to cutting the bonds between layers.  

In this work, we select $A$ and $B$ to be Pt and P, respectively, as Pt-P compounds are not common materials and also as P and N belong to the same group in the periodic table\textemdash the resulting single-layer structures should bear some similarity. According to the Materials Project,\cite{Jain2013} Pt and P can form only two stoichiometric compounds: PtP$_2$ and Pt$_2$P$_5$, both of which are the intermediate phases in Pt-P molten glasses.\cite{dahl1967crystal} Little research has been undertaken on these two compounds. Thomassen firstly determined the crystal structure of PtP$_2$ as the pyrite structure.\cite{Thomassen1929} Further work by Baghdadi and Thomas Schmidt {\it et al.} led to the accurate measurement of the crystal structure and single crystal prepared using a tin flux.\cite{Baghdadi1974,schmidt1990verfeinerung} We use this somewhat uncommon compound as an example of exploring the hidden Cairo tessellation in the pyrite structure to discover novel single-layer PtP$_2$ using DFT calculations. 
\section{Methods}
 We use the Vienna {\it Ab-initio} Simulation Package (VASP, version 5.4.4) for all of the DFT calculations.\cite{Kresse96p11169} We also use both the Perdew-Burke-Ernzerhof (PBE) and HSE06 hybrid density functionals\cite{Perdew96p3865, krukau2006influence} to approximate the exchange-correlation interactions.
 The electron-ion interactions are described by the PBE version of the potential dataset generated from the projector-augmented wave (PAW) method.\cite{Bloechl94p17953,Kresse99p1758} These potentials treat 5$d^9$6$s^1$ states of Pt atoms and 3$s^2$3$p^3$ states of P atoms as valence electrons. We adopt a surface slab model to simulate single-layer PtP$_2$. Each surface slab has a vacuum spacing of 18.0~\AA~ that is sufficiently large to separate the image interactions. We use the plane waves with their kinetic cutoff energy below 550~eV for approximating the total electron wave function. Moreover, we use $\Gamma$-centered $12~\times~12~\times~12$ and $12~\times~12~\times~1$ $k$-point grids for the integration in the reciprocal space for bulk and single-layer PtP$_2$, respectively.\cite{PhysRevB.13.5188} For the HSE06 calculations on bulk PtP$_2$, we use a $8~\times~8~\times~8$ $k$-point grid to reduce the computational time. We also decrease the $k$-point grid size for the calculations on single-layer PtP$_2$ to $8~\times~8~\times~8$ considering the spin-orbit coupling (SOC). All the HSE06 and SOC calculations are based on the optimized PBE structure.
 
We apply two methods to create an initial unit cell of single-layer PtP$_2$ for further geometry optimizations. In the first method (M1), we carve out a single-layer PtP$_2$ from its bulk structure. This initial structure is a buckled structure with non-coplanar Pt and P atoms. The second method (M2) is based on the first one, but we set all the atomic coordinates to be co-planar. By using these two methods, we expect to find two structures with local energy minima. The unit cells used in both methods consist of two formula units. VASP fully optimizes the in-plane lattice constants and atomic positions of the two unit cells of single-layer PtP$_2$ until the threshold (0.01 eV/\AA) of inter-atomic forces is reached. At each step of the geometry optimizations, we set the total energy convergence to 10$^{-6}$ eV.

We employ Phonopy\cite{phonopy} and VASP to obtain the phonon spectrum of single-layer PtP$_2$ following three steps. First, we use Phonopy to generate $3~\times~3\times 1$ supercells. Next, we use VASP to calculate the inter-atomic forces for each supercell using a $4~\times~4~\times~1$ $k$-point grid. Finally, the forces are collected and post-processed by Phonopy to compute phonon frequencies at each wave vector along a high-symmetry $k$-point path.
\section{Results and Discussion}
We begin with computing the lattice constant and the formation energy $E_\mathrm{f}^\mathrm{bulk}$ of bulk PtP$_2$. $E_\mathrm{f}^\mathrm{bulk}$ is calculated the as the difference between the total energy of bulk PtP$_2$ with reference to the energies of face-centered cubic Pt and monoclinic P. The calculated lattice constant and $E_\mathrm{f}^\mathrm{bulk}$ are 5.75~\AA~and -697 meV/atom, respectively, both agreeing well with previous results (5.76~\AA~and -692 meV/atom, respectively) of DFT calculations recorded in the Materials Project (Materials Project id: mp-730).\cite{Jain2013}  

\begin{figure}
 \includegraphics[width=8cm]{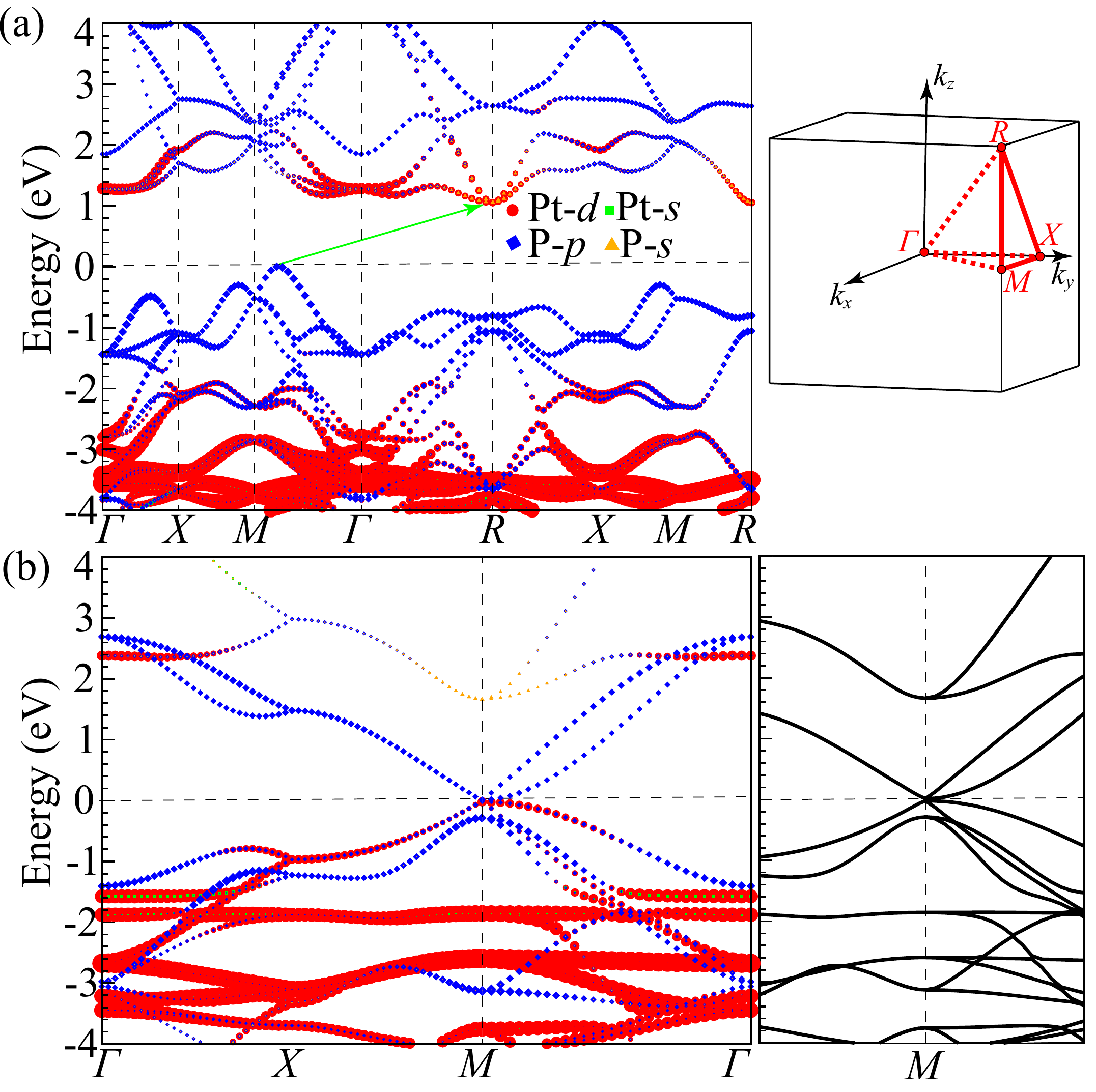}
  \caption{(a) Calculated orbital-resolved band structure of bulk PtP$_2$. The first Brillouin zone and high-symmetry $k$-point paths are shown in the right side panel. The green arrow is used to aid the view of the indirect band gap. (b) orbital-resolved band structure of single-layer PtP$_2$ with the Cairo tessellation. The right side panel shows the zoomed in band structure near the $M$ point. The valence band maximum in (a) is set to zero.}
  \label{fig:band}
\end{figure}

\begin{figure}
 \includegraphics[width=8cm]{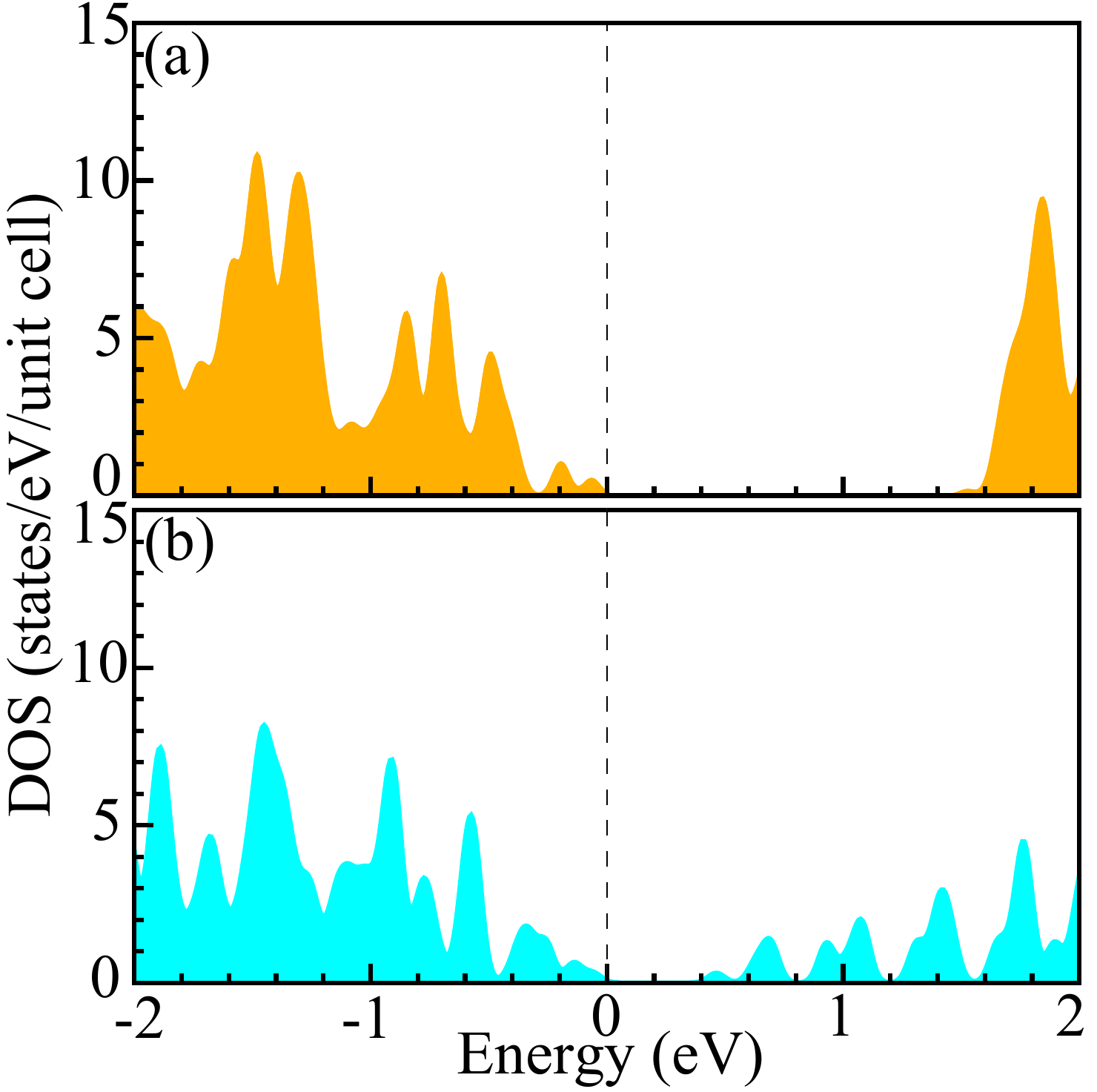}
  \caption{Density of states (DOS) of (a) bulk and (b) single-layer PtP$_2$ calculated using the HSE06 functional. The valence band maximum is set to zero.}
  \label{fig:hseband}
\end{figure}

We next calculate the band structure of bulk PtP$_2$. Figure~\ref{fig:band}(a) shows the theoretical orbital-resolved band structure of bulk PtP$_2$  computed with the PBE functional. Consistent again with the bandgap (1.02 eV) documented in the Materials Project,\cite{Jain2013} our calculated band structure shows that bulk PtP$_2$ is a semiconductor with an indirect PBE bandgap of 1.06 eV. The valence band maximum (VBM) is located at a $k$ point between the $M$ and $\Gamma$ points, and the conduction band minimum (CBM) occurs at the $R$ point. The $d$ orbitals of Pt atoms dominate the VBM, and the CBM originates from the contributions of $d$ orbitals of Pt and $s$ orbitals of P atoms. Because the PBE functional leads to underestimated bandgaps,\cite{PhysRevLett.51.1884} we further use the HSE06 hybrid density functional to correct the bandgap. Figure~\ref{fig:hseband} displays the HSE06 density of states of bulk PtP$_2$, showing that the corrected bandgap is 1.59 eV. This bandgap is within the visible light spectrum, so bulk PtP$_2$ may be useful for solar-energy conversion applications.

Having calculated the properties of bulk PtP$_2$, we set to focus on single-layer PtP$_2$. We mentioned in the method section that we use two different initial geometries to obtain the stable structure of single-layer PtP$_2$. In the M1 method, the initial out-of-plane distance distance $d$ \textemdash  between the sub planes of Pt and P atoms\textemdash is 0.63\AA ~determined by the bulk pyrite structure. But the geometry optimization transforms the structure into a nearly completely planar structure with a negligible $d$ of 0.003~\AA. In this method, the cross section of the surface slab also becomes slightly off a square. In the M2 method, the resulting structure is entirely planar and the cross section is strictly a square. The energy of the resulting structure from the M2 method is almost the same as that from the M1 method, but is trivially smaller by 0.1 meV per formula unit. We therefore conclude that single-layer PtP$2$ prefers adopting a fully planar structure. 

\begin{figure}
 \includegraphics[width=8cm]{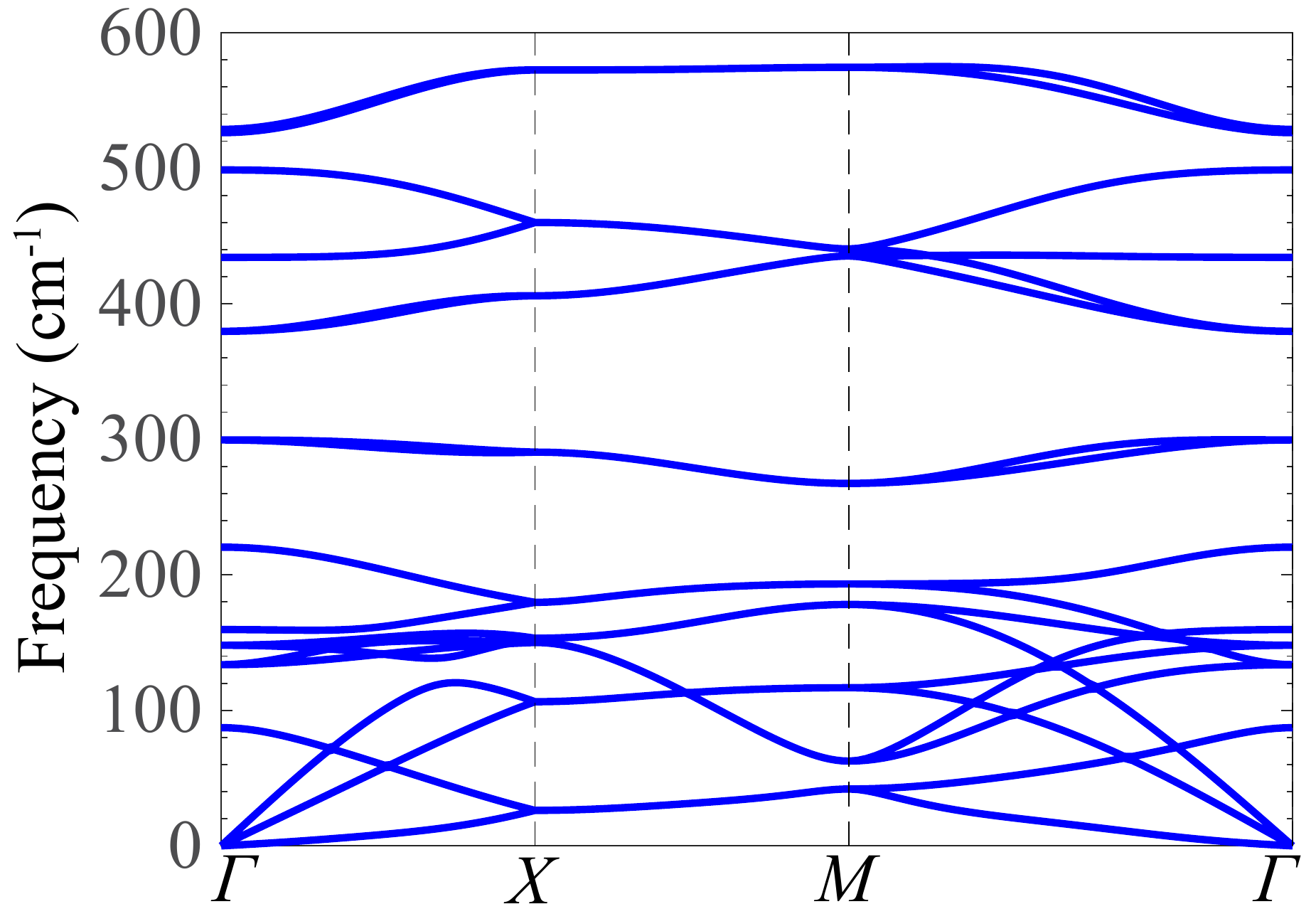}
  \caption{Predicted phonon spectrum of single-layer PtP$_2$ with the Cairo tessellation calculated at the DFT-PBE level of theory.}
  \label{fig:phonon}
\end{figure}

Figure 1(c) illustrates the side view of single-layer PtP$_2$ with the optimized, planar structure. The symmetry analysis performed by Phonopy shows that the space group of single-layer PtP$_2$ is P4/mbm (No.~127), corresponding to a lower symmetry than bulk PtP$_2$. As a result of the structure flattening, the calculated in-plane lattice constant (5.83~\AA) is slightly larger than that (5.75~\AA) of bulk PtP$_2$. To confirm that the single-layer PtP$_2$ with the planar, pentagonal structure is dynamically stable, we calculate the phonon spectrum, which is shown in Figure \ref{fig:phonon}. The absence of imaginary phonon modes corroborates the dynamical stability.

\begin{figure*}
 \includegraphics[width=16cm]{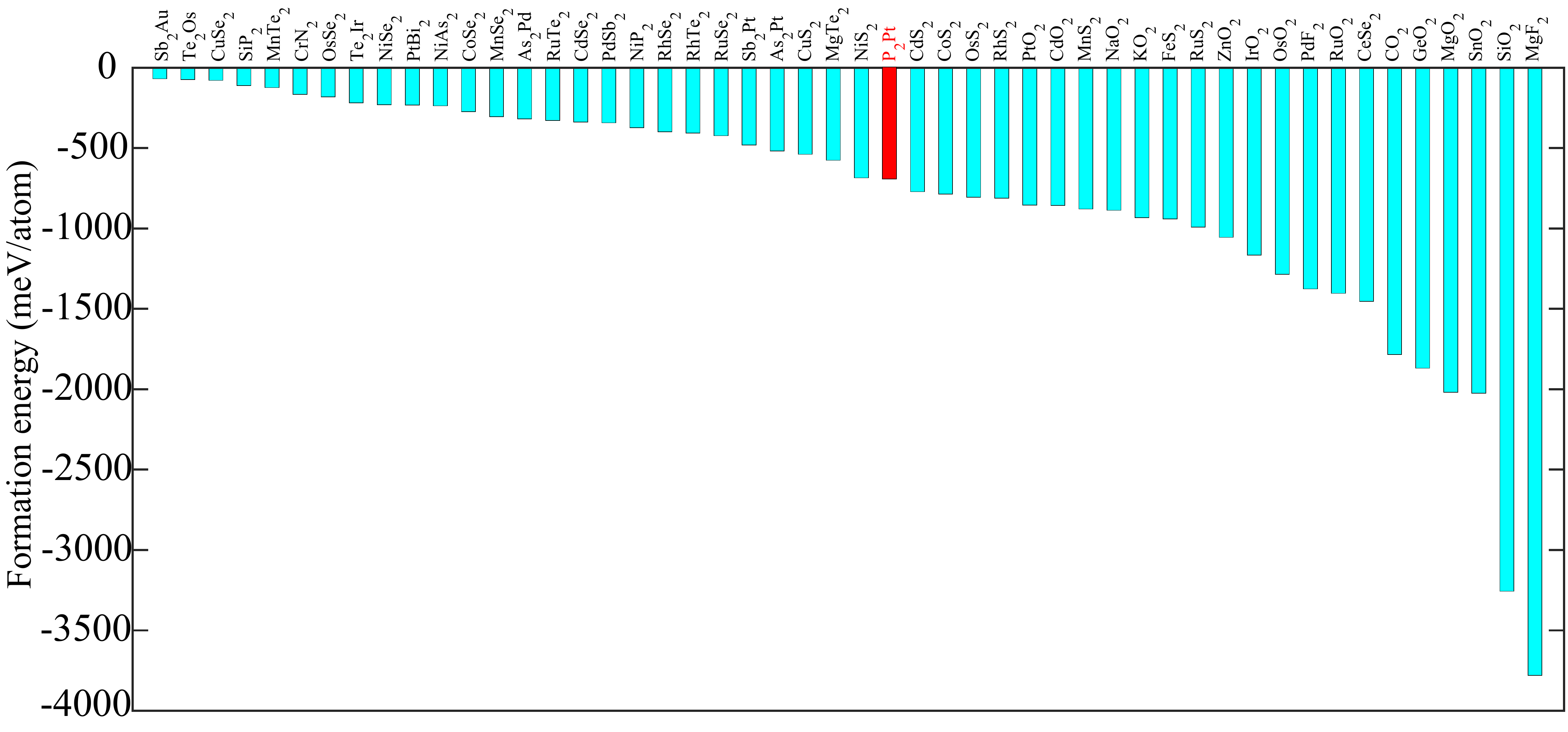}
  \caption{Bulk formation energies from data-mining the Materials Project\cite{Jain2013} for the compounds with the pyrite structure, a general chemical formula $AB_2$, and space group $pa\bar3$.}
  \label{fig:bar}
\end{figure*}

We then compare the geometry of a pentagon in the optimized single-layer PtP$_2$ structure to a type 2 pentagon illustrated in Fig.~\ref{fig:structure}(a). We calculate the optimized, nearest-neighboring Pt-P and P-P bond length as 2.30 and 2.08~\AA, respectively, corresponding to one constraint (b = c = d = e) of the special type 2 pentagon. The P-P-Pt, P-Pt-P, Pt-P-Pt bond angles are A = E = 116.4$^\circ$, B = D = 90$^\circ$, and C = 127.2$^\circ$, respectively. All of these side lengths and angles satisfy the minimum geometry constraints imposed on type 2 pentagon, {\it i.e.,} c = e and B + D = 180$^\circ$. Such a pentagonal geometry confirms that single-layer PtP$_2$ exhibits the Cairo tessellation.

To rule out the possibility of Pt and P forming a hexagonal single-layer structure, we also compute the energies of single-layer PtP$_2$ with the hexagonal trigonal prismatic (2$H$) and octahedral (1$T$) structures. We find that the energies of these two structures\textemdash both are found to be metallic\textemdash are higher than that of the pentagonal structure with the Cairo tessellation by 534 and 311 meV/atom, respectively. Namely, the stability of single-layer PtP$_2$ with the three structures follows the order of stability from the highest to the lowest: Pentagonal \textgreater~1$T$~\textgreater~2$H$. As such, the term single-layer PtP$_2$ henceforth refers to the the pentagonal structure with the Cairo tessellation. 

We then calculate the formation energy $E_\mathrm{f}^\mathrm{SL}$ of single-layer PtP$_2$ using the energy of its bulk counterpart as the reference.\cite{zhuang2014computational} We determine the $E_\mathrm{f}$ as 410 meV/atom, which is somewhat large, excluding the feasibility of obtaining single-layer PtP$_2$ from bulk PtP$_2$ via the mechanical exfoliation method as used to obtain graphene .\cite{singh2015computational} A most viable method to obtain single-layer PtP$_2$ is therefore via a chemical method such as the chemical vapor decomposition. 

We attempt to correlate the formation energies of bulk ($E_\mathrm{f}^\mathrm{bulk}$  = -697 meV/atom)  and single-layer PtP$_2$ ($E_\mathrm{f}^\mathrm{SL}$ = 410  meV/atom). We use a simplified model assuming that the nearest neighboring Pt-P and P-P bonds contribute the most significantly to the formation energies. We count the number of Pt-P and P-P bonds bonds in bulk PtP$_2$ as 24 and 4, respectively, for the 12 atoms in a unit cell. In other words, each bulk unit cell has 2 Pt-P and 1/3 P-P bonds per atom. Mathematically, we write
\begin{equation}
E_\mathrm{f}^\mathrm{bulk} = 2E_\mathrm{Pt-P} + 1/3 E_\mathrm{P-P}, 
\label{eq1}
\end{equation}
where $E_\mathrm{Pt-P}$ and $E_\mathrm{P-P}$ are the energies of the Pt-P and P-P bonds, respectively.  Similarly, the 6-atom unit cell of single-layer PtP$_2$ has 8 Pt-P and 2 P-P bonds, corresponding to 4/3 Pt-P and 1/3 P-P bonds per atom. Therefore, we have
\begin{equation}
E_\mathrm{f}^\mathrm{SL} = 4/3E_\mathrm{Pt-P} + 1/3E_\mathrm{P-P}. 
\label{eq2}
\end{equation}
Taking the difference of Eqs.\ref{eq1} and \ref{eq2} gives
\begin{equation}
E_\mathrm{f}^\mathrm{SL} = E_\mathrm{f}^\mathrm{bulk} - 2/3E_\mathrm{Pt-P}.
\label{eq3}
\end{equation}

Eqs.~\ref{eq1} and \ref{eq2} show that the number of P-P bonds remains the same when transforming from the bulk to single layer. Eq.~\ref{eq3} shows that the energy cost for the dimension reduction is  equivalent of removing 2/3 Pt-P bonds. Therefore the Pt-P bond energy is calculated as 615 meV/atom. This oversimplified bond-counting model seems to indicate that the smaller the bulk formation energy (smaller $E_\mathrm{f}^\mathrm{bulk}$) of a compound with the pyrite structure, the less energy-consuming (smaller $E_\mathrm{f}^\mathrm{SL}$) to obtain a single-layer pentagonal structure.

Following the above argument, we perform a data-mining operation in the Materials Project to identify all of the compounds with the pyrite structure, a general chemical formula $AB_2$, and space group $pa\bar3$. We find 50 such compounds with negative formation energies, implying they are stable in the bulk form. Figure~\ref{fig:bar} shows that the formation energies of these $AB_2$ compounds range widely from 69 meV/atom for AuSb$_2$ to 3780 meV/atom for MgF$_2$. We expect that the compounds whose bulk formation energies lie in the left hand side of the histogram correspond to small single-layer format energies, enhancing the possibility of obtaining the single-layer form of these compounds. We leave the calculations of the single-layer formation energies and characterization of these compounds as future work.     

\begin{figure}
 \includegraphics[width=8cm]{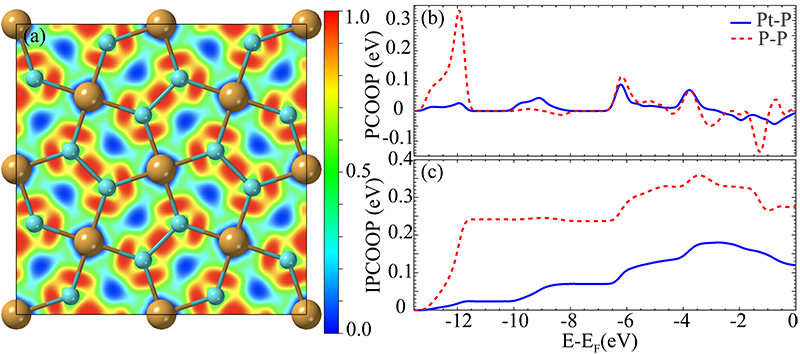}
  \caption{(a) The electron localization function of single-layer PtP$_2$. (b) Projected and (c) integrated crystal orbital overlap population (PCOOP and ICOOP) of the Pt-P and P-P bonds in single-layer PtP$_2$.}
  \label{fig:coop}
\end{figure}

We now focus on the electronic structure of single-layer PtP$_2$. We first calculate the Bader charge transfer to understand the bonding characteristics of Pt-P and P-P bonds.\cite{henkelman2006fast,tang2009grid} Consistent with the slightly more electronegativity of Pt than P (2.28 and 2.19 for Pt and P, respectively, in the Pauling scale\cite{pauling1945nature}), we find that 0.23 electrons are transferred from P to Pt in single-layer PtP$_2$, indicating the Pt-P bond is of the ionic nature. We next compute the electron localization functional (ELF).\cite{becke1990simple} The calculated ELF of single-layer PtP$_2$ is shown in Fig.\ref{fig:coop}(a). We see that the ELF values near the P atoms are almost equal to 1.0, showing that the electrons are localized around the P atoms in the Pt-P bond indicative of ionic bonding. The ELF results also show the electrons are shared in the P-P bond, suggesting covalent bonding. Both the Bader charge analysis and ELF show that single-layer Pt$P_2$ exhibits mixed types of ionic and covalent bonds. Furthermore, to provide a metric describing the strength of the Pt-P and P-P bonds, we calculate the crystal orbital overlap population (COOP) for the P-Pt and P-P bonds in a unit cell using the LOBSTER (Local Orbital Basis Suite Towards Electronic?Structure Reconstruction) tool.\cite{maintz2016lobster} Fig.~\ref{fig:coop} shows the projected COOP (PCOOP) as a function of the electron energy. We observe that both the Pt-P and P-P bonds exhibit bonding and antibonding characteristics \textemdash represented by positive and negative PCOOP, respectively \textemdash below the Fermi level. The integrated COOP (ICOOP) is shown in Fig.~\ref{fig:coop}(c). At the Fermi level, the ICOOP values for the Pt-P and P-P bonds are 0.12 and 0.27, respectively, showing that the P-P covalent bond is stronger than the ionic Pt-P bond.   

\begin{figure}
 \includegraphics[width=8cm]{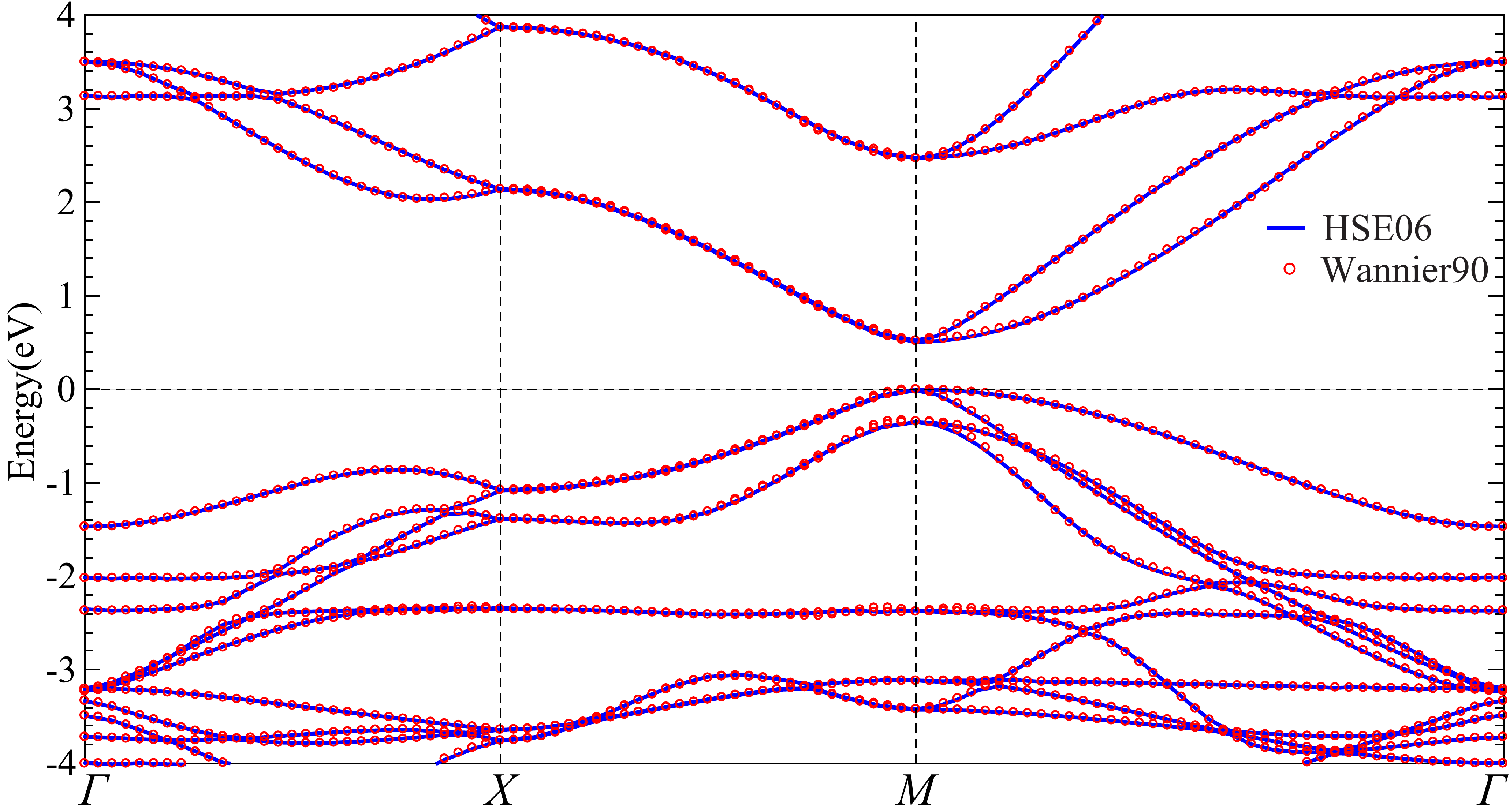}
  \caption{Band structure of single-layer PtP$_2$ calculated using the HSE06 functional and interpolated by the Wannier90 package. The valence band maximum is set to zero.}
  \label{fig:hsewf}
\end{figure}

Reducing bulk PtP$_2$ to single-layer nanosheets causes a drastic change in the band structure, as can be seen in Fig.~\ref{fig:band}(b). We observe a four-fold degeneracy of the conduction and valence bands at the $M$ point, showing the metallic behavior of single-layer PtP$_2$. Similar to the bulk band structure, the $d$ and $s$ orbitals of Pt atoms and the $p$ and $s$ orbitals of P atoms all contribute to form the band structure of single-layer PtP$_2$. But the transition from a semiconducting bulk to a metallic single layer seems surprising to some extent. We mentioned that the PBE functional is well known to underestimate bandgap, which may also lead to an incorrect conclusion that a semiconductor with a narrow bandgap is considered to be metallic. We therefore use the more accurate HSE06 hybrid functional to calculate the density of states to confirm whether single-layer PtP$_2$ is truly metallic. Figure~\ref{fig:hseband}(b) shows that the computed DOS using the HSE06 functional exhibits a bandgap of 0.52 eV. The corresponding band structure shown in Figure~\ref{fig:hsewf} further reveals that the bandgap is also a direct bandgap with the CBM and VBM both at the $M$ point. Although the bandgap of single-layer PtP$_2$ is narrower than that of bulk PtP$_2$, semiconducting single-layer PtP$_2$ with a direct bandgap may be useful in applications such as infrared detectors.\cite{mcgill1993prospects} 

Several narrow-bandgap, single-layer semiconductors such as 1$T$' MoS$_2$\cite{qian2014quantum} and PbTe \cite{liu2015crystal} have been predicted to be topological insulators, where the bulk states behave as an insulator but the surface states show a spin-helical Dirac cone.\cite{RevModPhys.82.3045,RevModPhys.83.1057} To examine the possible existence of such a topological phase in single-layer PtP$_2$, we study the surface states of single-layer PtP$_2$. We use tight-binding-like Wannier parameters and the iterative Green's function method\cite{PhysRevB.28.4397, sancho1984quick} as implemented in the WannierTools package\cite{WU2018405} to compute the surface states. According to the orbital-resolved band structure shown in Fig.~\ref{fig:band}(b), we use the Wannier90 package (version 1.2) to obtain 28 Wannier orbitals ($s$ and $d$ orbitals of two Pt atoms and $s$ and $p$ orbitals of four P atoms in a unit cell) projected from converged HSE06 wave functions in VASP calculations. Correspondingly, 56 Wannier orbitals are used if the spin-orbit coupling (SOC) is taken into account. As can be seen from Fig.\ref{fig:hsewf}, the obtained Wannier parameters are accurate enough to reproduce the HSE06 band structure. Without considering the SOC effect, we observe two degenerate surface bands near the bandgap. Including the SOC, the degeneracy is lifted, doubling the number of surface bands. But the spin-helical Dirac states remain absent, excluding single-layer PtP$_2$ as a topological insulator, possibly due to the weak SOC. 
\begin{figure}
 \includegraphics[width=8cm]{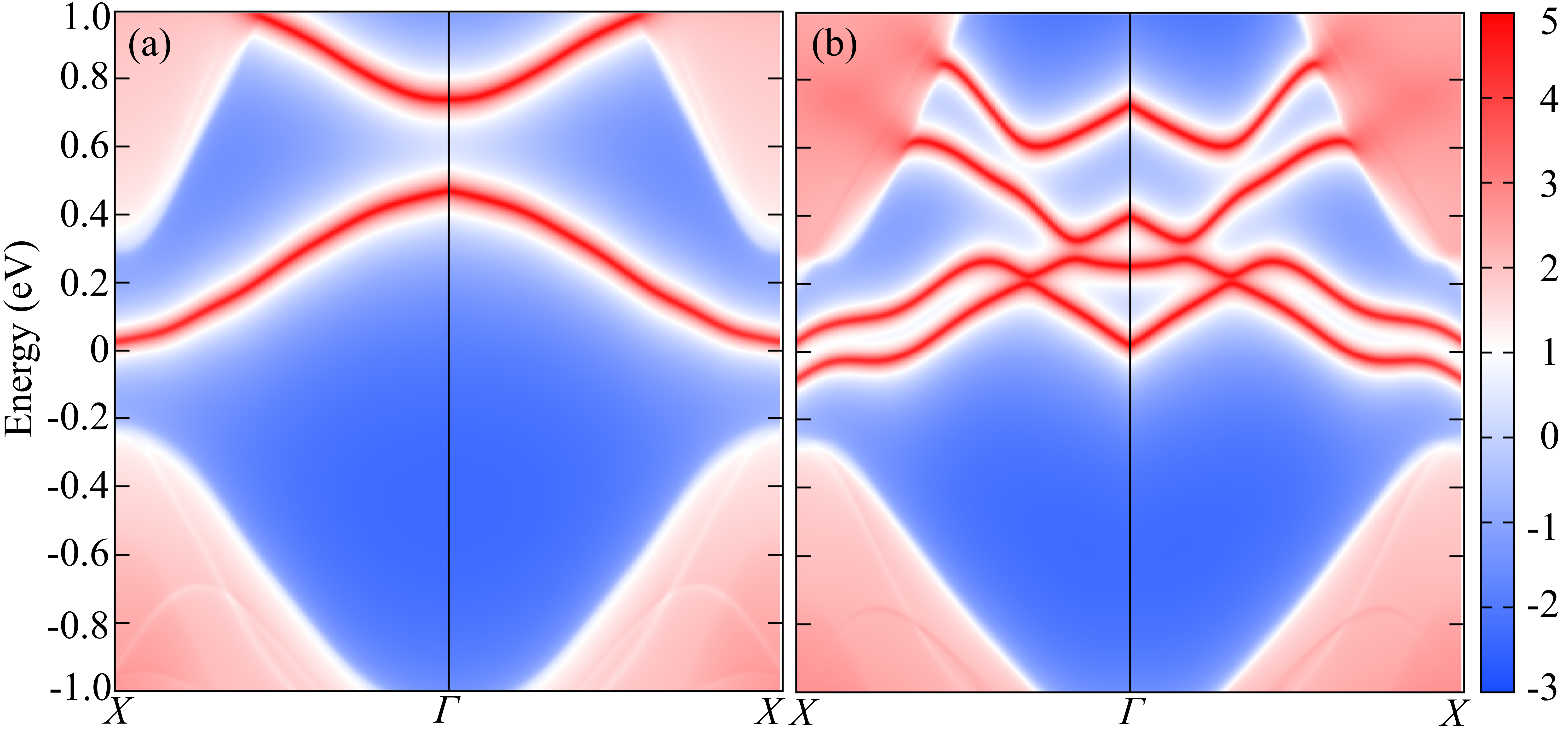}
  \caption{Surface band structures of single-layer PtP$_2$ (a) without and (b) with spin-orbit coupling.}
  \label{fig:surfs}
\end{figure}
\section{Conclusions}
In conclusion, we demonstrated an example of discovering 2D materials by uncovering the hidden Cairo tessellation in the pyrite structure. We predicted that single-layer PtP$_2$ is a new 2D material with a fully planar, pentagonal structure. We found that the PBE functional incorrectly predicted single-layer PtP$_2$ to be metallic. But the more accurate HSE06 hybrid density functional calculations showed that single-layer PtP$_2$ indeed exhibits a reduced, direct bandgap in comparison with bulk PtP$_2$. A future work could be integrating the procedure of computational characterization followed in this work into a high-throughput framework for accelerating discovery of pentagonal 2D materials. 
\begin{acknowledgments}
We thank the start-up funds from Arizona State University. We also thank Drs. Jianfeng Wang, Bing Huang, and Xiaofeng Qian for helpful discussions. This research used computational resources of the Texas Advanced Computing Center under Contracts No.TG-DMR170070. 
\end{acknowledgments}
\bibliography{references}
\end{document}